\documentclass[aps,twocolumn,reprint]{revtex4-1}

\usepackage{graphicx}
\usepackage[english]{babel}
\usepackage{amsmath, amssymb, amsfonts}
\usepackage{bm}
\usepackage{color}

\newlength{\figwidth}
\newlength{\lfig}
\newlength{\sfig}
\newcommand{\RbCs}{$^{87}$Rb$^{133}$Cs}
\newcommand{\KCs}{$^{39}$K$^{133}$Cs}
\newcommand{\CaF}{$^{40}$Ca$^{19}$F}

\begin{document}
\title{Microwave shielding of ultracold polar molecules}
\date{\today}
\author{Tijs Karman}
\author{Jeremy M. Hutson}
\affiliation{Joint Quantum Centre (JQC) Durham-Newcastle, Department of
Chemistry, Durham University, South Road, Durham, DH1 3LE, United Kingdom}

\begin{abstract}
We use microwaves to engineer repulsive long-range interactions between
ultracold polar molecules. The resulting shielding suppresses various loss
mechanisms and provides large elastic cross sections. Hyperfine interactions
limit the shielding under realistic conditions, but a magnetic field allows
suppression of the losses to below $10^{-14}$~cm$^3$~s$^{-1}$. The mechanism
and optimum conditions for shielding differ substantially from those proposed
by Gorshkov et al.\ [Phys.\ Rev.\ Lett.\ 101, 073201 (2008)], and do not
require cancelation of the long-range dipole-dipole interaction that is vital
to many applications.
\end{abstract}

\maketitle

A variety of polar molecules have now been produced at \cite{Ni:KRb:2008,
Takekoshi:RbCs:2014, Molony:RbCs:2014, Park:NaK:2015, Guo:NaRb:2016,
Rvachov:2017}, or cooled down to \cite{cheng:16, Truppe:MOT:2017,
McCarron:2018}, ultracold temperatures. Potential applications include quantum
simulation \cite{Santos:2000, Baranov:2012}, quantum computing
\cite{DeMille:2002, Yelin:2006} and the creation of novel quantum phases
\cite{Buechler:2007, Lechner:2013}. All these applications require high
densities, where collisional losses becomes important. Even chemically stable
molecules in their absolute ground state, which possess no two-body loss
mechanisms, may undergo short-range three-body loss that is amplified by
long-lived two-body collisions \cite{mayle:12,mayle:13,Ye:2018}. Short-range
losses have been suppressed experimentally for fermionic molecules by a
combination of strong electric fields and confinement \cite{demiranda:11}.
However, this approach is not feasible for bosons \cite{Quemener:2011}. In this
paper, we use microwaves to engineer repulsive long-range interactions that
shield molecular collisions.  Our approach does not require confinement to 2
dimensions as in Refs.\ \cite{Buechler:2007, micheli:07}, and can be applied to
both bosonic and fermionic species.

\begin{figure}
\begin{center}
\includegraphics[width=\figwidth,clip]{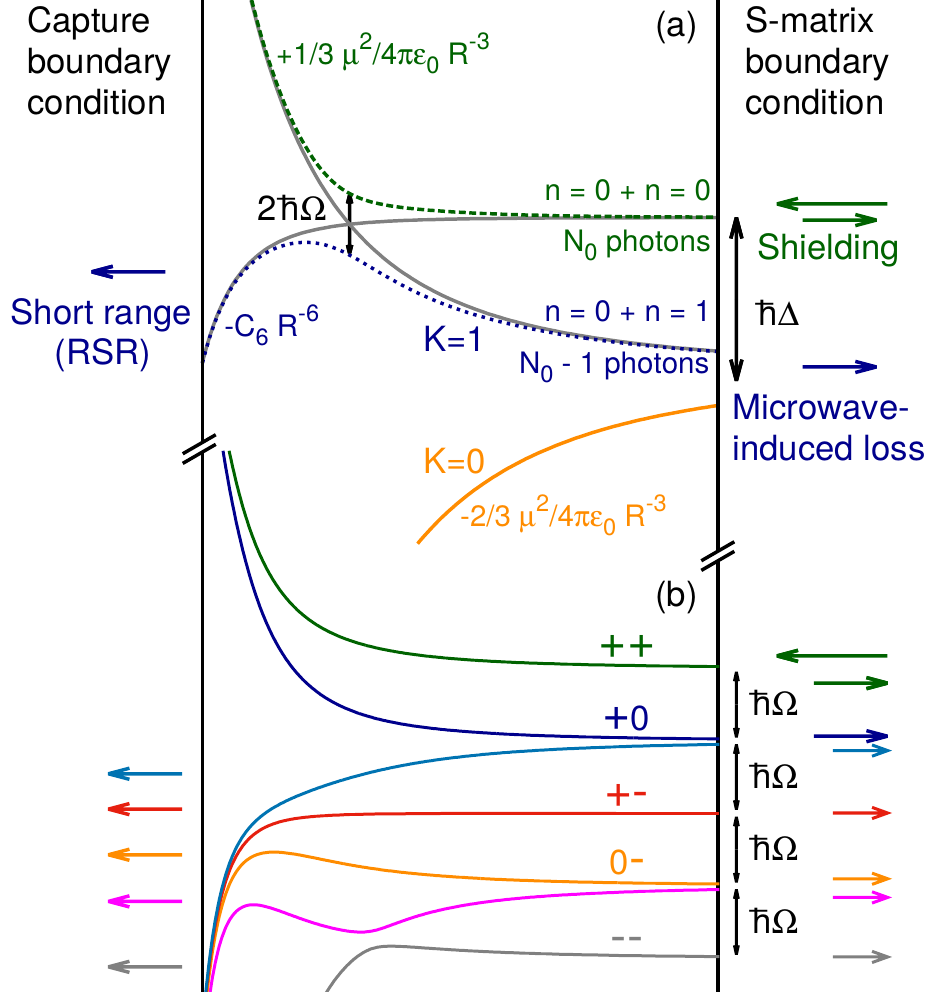}
\caption{ \label{fig:schematic}
Schematic representation of the potential curves relevant to microwave shielding.
Panels (a) and (b) correspond to $\Omega \ll \Delta$ and $\Omega\gg\Delta$, respectively.
The boundary conditions imposed in the coupled-channels calculations are indicated.
Green arrows indicate incoming and elastically scattered flux,
whereas the remaining arrows on the right- and left-hand sides indicate microwave-induced loss and reaching short range (RSR), respectively.
}
\end{center}
\end{figure}

Fig.~\ref{fig:schematic}(a) shows the shielding mechanism schematically in the
low-intensity limit. Microwave radiation is blue detuned by $\Delta$ from the
$n=0\rightarrow1$ rotational transition of the molecule. The field-dressed
state with one molecule rotationally excited $(n=1)$ is energetically below the
bare state with both molecules in the ground state $(n=0)$ by $\hbar
\Delta$. The resonant dipole-dipole interaction splits the lower threshold into
repulsive $|K|=1$ and attractive $K=0$ states. Here, $K$ is the projection of
the rotational angular momentum onto the intermolecular axis, which is a good
quantum number when Coriolis and field-dependent couplings are neglected. The repulsive $K=1$ states
cross the bare ground state at the Condon point, which moves inwards as
$\Delta$ increases. This crossing is avoided by $2\hbar\Omega$, where $\Omega$
is the Rabi frequency. The upper adiabatic curve is repulsive and provides
shielding. This is closely analogous to optical blue-shielding for atoms
\cite{weiner:99}.

Microwave-dressed molecules typically have weaker resonant dipole interactions than optically dressed atoms and
need larger values of $\Omega$ for optimum shielding. For high intensities the
individual monomer states are even and odd linear combinations $|\pm\rangle$ of
the field-dressed states $|g\rangle = |0,0,0\rangle$ and $|g\rangle =
|1,1,-1\rangle$, with energies $\pm\hbar\Omega$. In the ket $|n,m_n,N\rangle$,
$N$ is the number of photons of $\sigma^+$ polarization and $m_n$ is the
projection of $n$ onto the microwave propagation axis. There are also dark
states $|0\rangle$ corresponding to $|1,0,-1\rangle$ and $|1,-1,-1\rangle$.
This produces 5 thresholds separated by approximately $\hbar\Omega$, as shown
in Fig.~\ref{fig:schematic}(b). The top adiabatic curve is again repulsive and
provides shielding.

Our goal is to find conditions, $\Omega$ and $\Delta$, under which the
collision dynamics is adiabatic and follows the repulsive shielding potential.
We calculate the potential curves and couplings from a Hamiltonian that
describes the molecules as rigid rotors interacting by dipole-dipole
interactions. It also includes end-over-end rotation of the molecular pair (not
included above) and interactions with electric, magnetic and microwave fields,
with hyperfine interactions where appropriate. We use a basis set consisting of
symmetrized products of spherical harmonics for the rotation of both molecules
and the end-over-end rotation, as well as electron and nuclear spin states. A
full description of the Hamiltonian and examples of the resulting adiabatic
curves are given in the supplemental material Sec.~S1.

We perform numerically exact coupled-channels scattering calculations of two
different types of loss. The coupled-channels approach is essential, because
semiclassical approximations such as Landau-Zener break down when the
wavelength is large compared to the width of the crossing. We propagate two
sets of linearly independent solutions of the coupled-channels equations, using
the renormalized Numerov method \cite{Johnson:1978}, and apply both capture
boundary conditions at short range and $S$-matrix boundary conditions at long
range \cite{light:76,clary:87,janssen:13}. We calculate the probability of
reaching short range (RSR) and the corresponding rate coefficient. There is
evidence that flux that reaches short range is lost with high probability, even
for non-reactive molecules \cite{Ye:2018}. In addition, some of the reflected
flux is lost, for example by absorbing a microwave photon, accompanied by
kinetic energy release. We also calculate the probabilities and rates for this
\emph{microwave-induced} loss. These two types of loss are illustrated in
Fig.~\ref{fig:schematic}. The remaining flux is shielded and scatters
elastically.

\begin{figure}
\begin{center}
\includegraphics[width=\figwidth,clip]{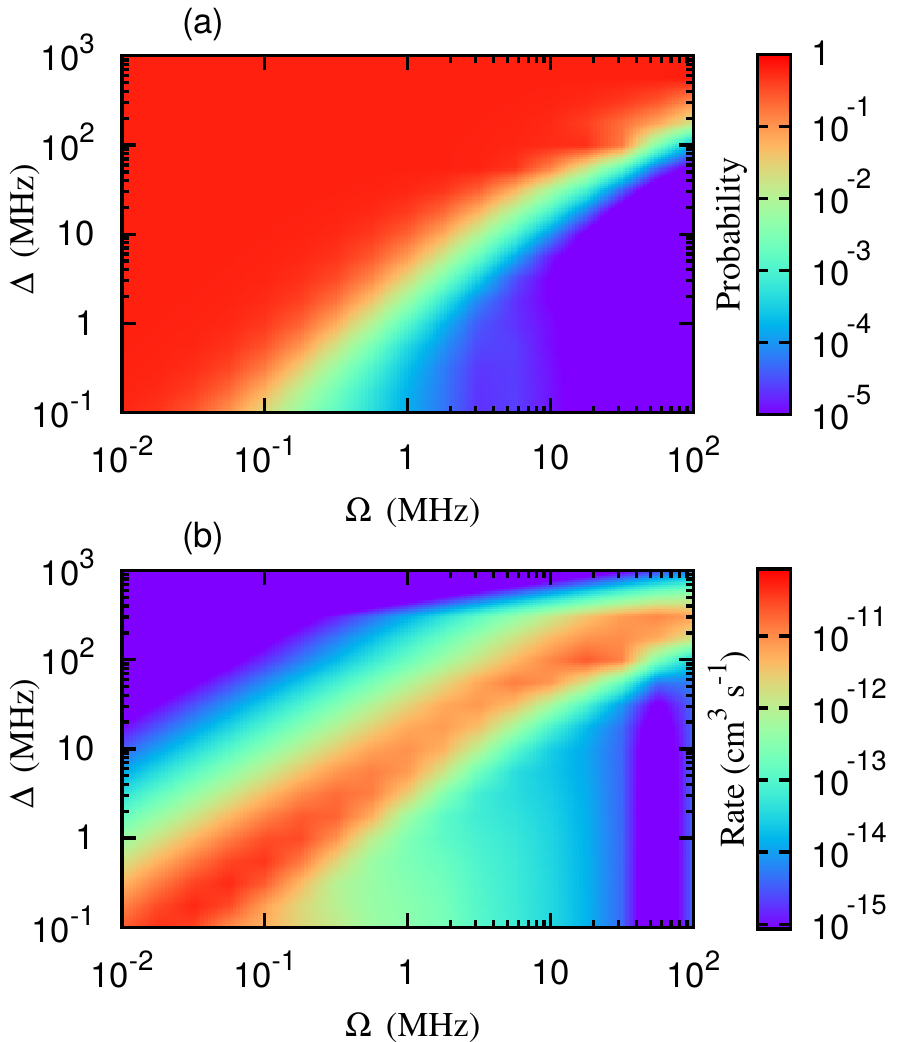}
\caption{ \label{fig:shield}
Probability for RSR (a) and microwave-induced loss rate (b), as a function of
$\Delta$ and $\Omega$, for RbCs+RbCs collisions in circularly polarized
microwaves, without static fields or hyperfine interactions. The color codings
for probability and loss rate are equivalent and can be used to read either
panel. }
\end{center}
\end{figure}

Figure~\ref{fig:shield} shows the probabilities and rates for RSR and
microwave-induced loss as a function of $\Delta$ and $\Omega$. This calculation
is for RbCs+RbCs collisions in the presence of circularly polarized
($\sigma^+$) microwaves, without static fields or hyperfine interactions. For
large $\Omega$ and comparable or smaller $\Delta$, the probabilities for both
RSR and microwave-induced loss are small, indicating that shielding is
effective. Loss rates below $10^{-14}$~cm$^3$~s$^{-1}$ can be achieved for
feasible values of $\Omega$; such rates are low enough to allow lifetimes of
several seconds at densities that are high enough for Bose-Einstein
condensation. Shielding is ineffective for linearly polarized microwaves, as
shown in the supplemental material Sec.~S1.

Microwave shielding of polar molecules is ineffective for $\Omega\ll\Delta$.
This contrasts with blue-shielding for ultracold atoms, and arises both because
of the smaller transition dipoles for typical molecules and because of the
strong rotational dispersion interaction. For $\Omega\gtrsim\Delta$ there is
significant state mixing even for the separated molecules, and the molecules
must be prepared in the upper field-dressed state. This may be done either by
forming molecules directly in the upper state by STIRAP or by switching on the
microwave field adiabatically. For a linear intensity ramp, switching on the
microwaves over $1$~ms for $\Omega=10$~MHz and $\Delta=1$~MHz retains 99\% in
the upper adiabatic state, as described in the supplemental material
Sec.~S1. Considerably shorter times may be achieved with ramps that are
slower at low intensity.

For ultracold collisions, the strong dependence of the scattering length on the
position of the least-bound state usually precludes \emph{ab initio}
calculation of elastic cross sections $\sigma_{\rm el}$ \cite{Wallis:MgNH:2009,
Wallis:LiNH:2011}. In the presence of shielding, however, the molecules never
experience the inaccurately known short-range interactions, and the calculated
$\sigma_{\rm el}$ is quantitatively predictive. For RbCs molecules, shielded as
above with $\Delta=1$~MHz and $\Omega=10$~MHz, we obtain $\sigma_{\rm
el}=3.6\times 10^{-10}$~cm$^{2}$. This is large compared to the typical value
expected for unshielded RbCs molecules, which is $4\pi \bar{a}^2 = 1.8\times
10^{-11}$~cm$^{2}$. Here $\bar{a}$ is the mean scattering length
\cite{Gribakin:1993} that accounts for the rotational dispersion interaction.
The combination of large elastic and suppressed inelastic cross sections may
allow evaporative cooling of microwave-shielded polar molecules.

\begin{figure}
\begin{center}
\includegraphics[width=\figwidth,clip]{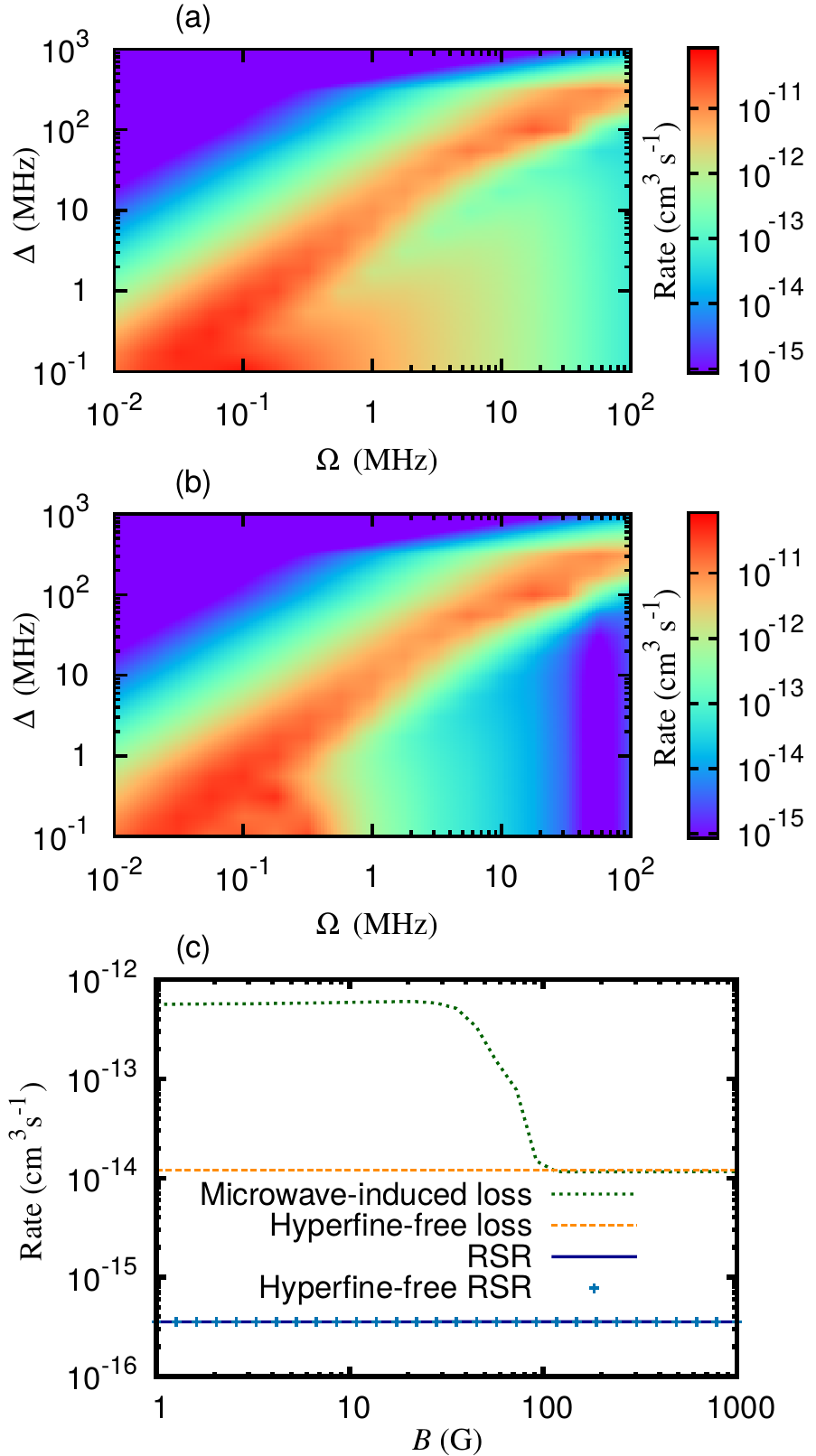}
\caption{ \label{fig:RbCs}
Shielding of collisions of RbCs molecules in the spin-stretched state by
circularly polarized microwaves, including hyperfine interactions. Panels (a)
and (b) show the microwave-induced loss rate in 0~G and 200~G magnetic fields,
respectively. Panel (c) shows the dependence of the RSR and
microwave-induced loss rates on magnetic field for fixed $\Omega=20$~MHz and
$\Delta=1$~MHz. }
\end{center}
\end{figure}

We next consider the effect of hyperfine interactions. These can cause losses
for molecules that are not present for atoms, because atomic hyperfine
splittings are much larger than $\Omega$ and $\Delta$. We carry out
coupled-channel calculations in a full basis set including nuclear spin
functions \cite{Aldegunde:polar:2008}. We initially consider
$^{87}$Rb$^{133}$Cs molecules in the spin-stretched $f=5, m_f=5$ state for
$n=0$, which can be produced and trapped experimentally
\cite{Takekoshi:RbCs:2014, Molony:RbCs:2014}. This state has the advantage that
there is only one allowed microwave transition for $\sigma^+$ polarization, to
the spin-stretched $f=6, m_f=6$ rotationally excited $n=1$ state
\cite{Gregory:RbCs-microwave:2016}. At low magnetic fields, the additional
channels resulting from hyperfine coupling produce greater microwave-induced
loss, as can be seen in Fig.~\ref{fig:RbCs}(a). However, a magnetic field of
200~G parallel to the microwave propagation axis recovers the effective
shielding obtained in the hyperfine-free case, as shown in
Fig.~\ref{fig:RbCs}(b). The transition between the low-field and high-field
regimes is shown in Fig.~\ref{fig:RbCs}(c) for fixed $\Omega=20$~MHz and
$\Delta=1$~MHz. The rate for RSR is small, as in the hyperfine-free case.

\begin{figure}
\begin{center}
\includegraphics[width=\figwidth,clip]{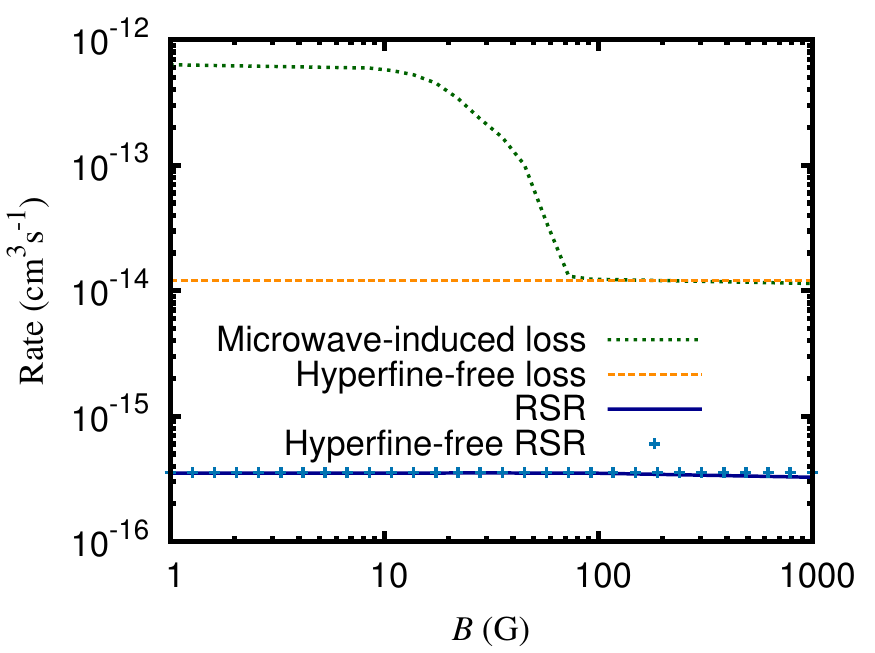}
\caption{ \label{fig:nonstretched}
The dependence of the microwave-induced loss on magnetic field for fixed
$\Omega=20$~MHz and $\Delta=1$~MHz, for collisions of RbCs molecules in the
non-spin-stretched $f=4, m_f=4$ state, including hyperfine interactions.}
\end{center}
\end{figure}

The spin-stretched state becomes the absolute ground state at magnetic fields
above 90~G. However, this is not a necessary or a sufficient condition for
effective shielding. Figure \ref{fig:nonstretched} shows the microwave-induced
loss for the non-spin-stretched $f=4,m_f=4$ state of RbCs as a function of
magnetic field, for $\Omega=20$~MHz and $\Delta=1$~MHz. The loss reduces to the
hyperfine-free value over much the same range of magnetic fields as for the
spin-stretched state. The $f=4,m_f=4$ state is not the absolute ground state at
any field; the suppression occurs because $m_n$ becomes a nearly good quantum
number at high fields. Microwave shielding may be achieved even for states that
are not spin-stretched and are not the absolute ground state.

Similar or better shielding should be achievable for other polar bialkali
molecules, where the hyperfine interactions are typically weaker than for RbCs
\cite{aldegunde:17}. The supplemental material Sec.~S1 gives results for
the case of $^{39}$K$^{133}$Cs, where the hyperfine couplings are weak enough
that substantial shielding can be achieved even in zero magnetic field. The
supplemental material also considers the $^2\Sigma$ molecule CaF, where
shielding is still effective but requires larger Rabi frequencies because of
stronger couplings involving the unpaired electron spin.

\begin{figure}
\begin{center}
\includegraphics[width=\figwidth,clip]{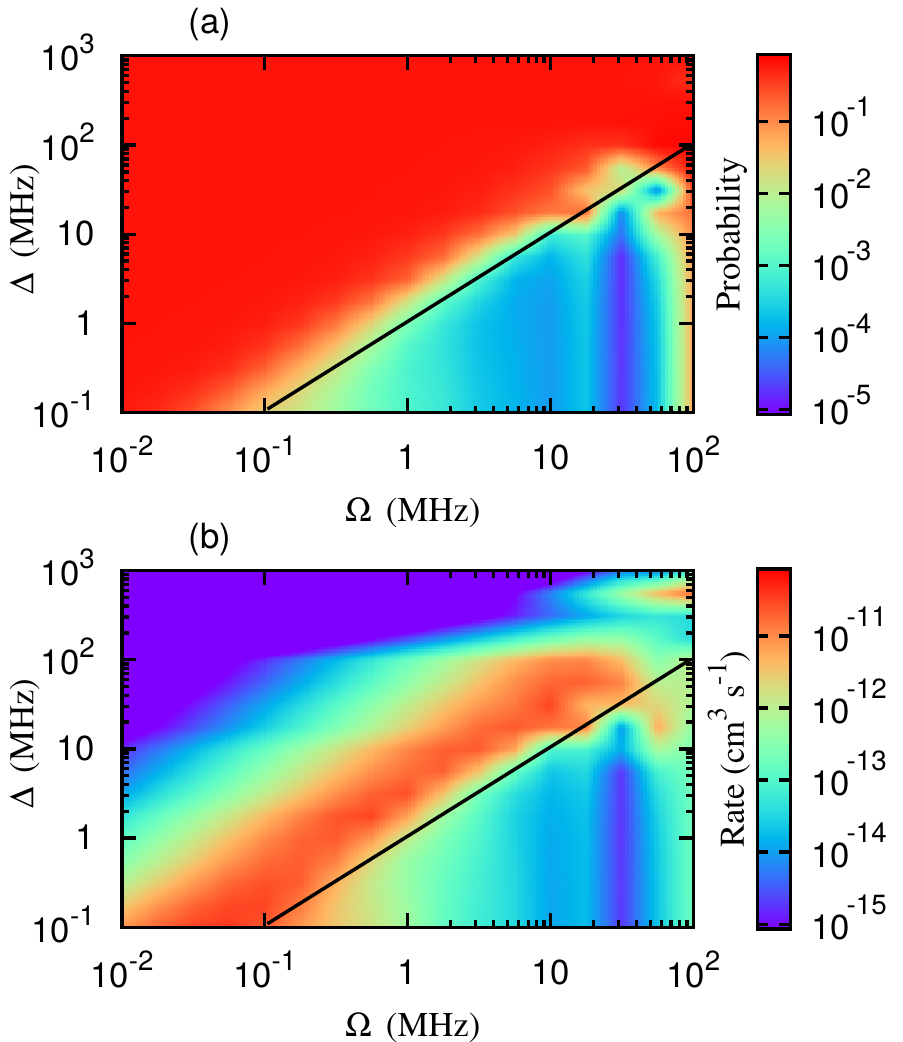}
\caption{ \label{fig:Efield}
Probability for RSR (a) and microwave-induced loss rate (b), as a function of
$\Delta$ and $\Omega$, for RbCs+RbCs collisions in circularly polarized
microwaves and an electric field $b_\mathrm{rot}/\mu$. The black lines
indicate $\Omega/\Delta$ chosen so that the microwave-induced and field-induced
dipole-dipole interactions cancel \cite{gorshkov:08}. The optimum conditions for
shielding are far from this line.}
\end{center}
\end{figure}

Gorshkov \emph{et al.}\ \cite{gorshkov:08} proposed a different mechanism for
microwave shielding in the presence of a static electric field. For a given
electric field, they chose $\Omega$ and $\Delta$ to cancel the first-order
dipole-dipole interaction. The dipole-dipole coupling then acts in second
order, producing an $R^{-6}$ interaction that is always repulsive for the upper
adiabatic state. They estimated loss rates using a semiclassical model of the
nonadiabatic transitions. We have calculated RSR probabilities and
microwave-induced loss rates for RbCs at an electric field of 0.9~kV/cm, which
optimizes the repulsive $R^{-6}$ shield \cite{gorshkov:08}. Our
coupled-channels results are shown in Fig.~\ref{fig:Efield}. The black line at
$\Omega/\Delta=0.95$ shows where cancelation of the first-order interaction
occurs. For this particular electric field, shielding starts to be effective at
values of $\Omega$ close to the line, but this is coincidental and is not true
for other electric fields. The optimum shielding is obtained for values of
$\Omega$ and $\Delta$ that are far from the line. It occurs at much higher
values of $\Omega/\Delta$, where there is no cancelation of the dipole-dipole
interaction. Microwave shielding can thus be realized in the presence of
first-order dipole-dipole interactions, which play an essential role in most
applications of ultracold polar molecules.

In conclusion, we have shown that collisions of ultracold polar molecules can
be shielded by circularly polarized microwave radiation tuned close to a
rotational transition. The microwaves prevent the collisions sampling the
short-range region, where both 2-body and 3-body loss processes may occur. We
have shown that hyperfine interactions may increase loss rates, but that this
can be suppressed in a magnetic field. Loss rates can be suppressed to below
$10^{-14}$~cm$^3$~s$^{-1}$, permitting lifetimes of seconds at densities
sufficient for Bose-Einstein condensation. Shielding also produces large
elastic cross sections, which combined with suppressed inelastic cross sections
may allow evaporative cooling. Shielding is also effective in external electric
fields, but the optimum parameters differ substantially from those proposed by
Gorshkov \emph{et al.}\ \cite{gorshkov:08} and do not require cancelation of
the field-induced dipole-dipole interaction.

\section*{Acknowledgement}
This work was supported by the U.K. Engineering and Physical Sciences Research
Council (EPSRC) Grants No.\ EP/P008275/1, EP/N007085/1 and EP/P01058X/1.

\noindent \emph{Note added:} After submission of this letter, we became aware
of parallel work by Lassabli\`ere and Qu\'em\'ener \cite{Lassabliere:2018} that
also considers the effect of microwave radiation on molecule-molecule
scattering lengths.

\widetext
\clearpage
\begin{center}
\textbf{\large S1. Supplement to: Microwave shielding of ultracold polar molecules}
\end{center}
\setcounter{equation}{0}
\setcounter{figure}{0}
\setcounter{table}{0}
\setcounter{page}{1}
\makeatletter
\renewcommand{\theequation}{S\arabic{equation}}
\renewcommand{\thefigure}{S\arabic{figure}}

In this supplemental material, we give the definition of the Hamiltonian used,
including the values of all parameters for the molecules considered. We define
the channel basis used in the scattering calculations, present adiabatic
potential curves, and discuss the calculation of cross sections, rates, and
probabilities for reaching short range (RSR) and microwave-induced loss. We
present further numerical results for the dependence on the microwave
polarization, as well as for shielding of \KCs\ and \CaF\ molecules.
Microwave-induced losses are weaker for KCs than for \RbCs, considered in the
main text. This illustrates that shielding of RbCs, which is shown to be
feasible, is a stringent test owing to its strong hyperfine interactions. For
CaF($^2\Sigma)$, microwave-induced loss is much larger due to the stronger
couplings involving the unpaired electron spin, but can still be suppressed in
a magnetic field. Finally, we calculate microwave switching times for state
preparation.

\section{Scattering calculations}

\subsection{Monomer Hamiltonian}

The molecules are modelled as rigid rotors with a dipole moment.
The monomer Hamiltonian of molecule $X$ is thus given by
\begin{align}
\hat{H}^{(X)} = b_\mathrm{rot} \hat{n}^2- \hat{\mu}^{(X)} \cdot \vec{E}_\mathrm{static} + \hat{H}_\mathrm{ac}^{(X)} + \hat{H}_\mathrm{hyperfine}^{(X)}.
\label{eq:monH}
\end{align}
The first term describes the rotational kinetic energy, with rotational constant $b_\mathrm{rot}$.
The second term describes the Stark interaction with a static electric field, $\vec{E}_\mathrm{static}$.
The third term above represents the interaction with a microwave electric field\cite{Cohen-Tannoudji:API:1998}
\begin{align}
\hat{H}_\mathrm{ac}^{(X)} = -\frac{E_\mathrm{ac}}{\sqrt{N_0}} \left[ \hat{\mu}_\sigma^{(X)} \hat{a}_\sigma + \hat{\mu}_\sigma^{(X)\dagger} \hat{a}_\sigma^\dagger\right] + \hbar\omega  \hat{a}_\sigma^\dagger \hat{a}_\sigma.
\end{align}
Here, $a_\sigma^\dagger$ and $a_\sigma$ are creation and annihilation operators
for photons with polarization $\sigma$ and angular frequency $\omega$. The microwave
electric field strength is given by $E_\mathrm{ac}$, and $N_0$ is the reference
number of photons. The dipole operator has spherical components $\sigma=0,\pm1$
which are related to the Cartesian components by $\hat{\mu}^{(X)}_0 =
\hat{\mu}^{(X)}_z$ and $\hat{\mu}^{(X)}_{\pm 1} = \mp \left( \hat{\mu}^{(X)}_x
\pm i \hat{\mu}^{(X)}_y\right)/\sqrt{2}$, corresponding to polarizations $\pi$
and $\sigma^\pm$.

The fourth term in Eq.~\eqref{eq:monH} represents the hyperfine Hamiltonian
\cite{aldegunde:08},
\begin{align}
\hat{H}_\mathrm{hyperfine}^{(X)} = \hat{H}_{eQq}^{(X_1)} + \hat{H}_{eQq}^{(X_2)} + c_1 \hat{i}^{(X_1)} \cdot \hat{n} + c_2 \hat{i}^{(X_2)} \cdot \hat{n} + H^{(X)}_3 + c_4 \hat{i}^{(X_1)} \cdot \hat{i}^{(X_2)} + \hat{H}^{(X)}_\mathrm{Zeeman}.
\end{align}
These describe various interactions that involve the spins of the nuclei $X_1$
and $X_2$ of molecule $X$.

For nuclear spins $i^{(x)}>1/2$, the dominant
interaction is that between the nuclear quadrupole moment and the internal
electric field gradient. This can be written as
\begin{align}
\hat{H}_{eQq}^{(x)} = ({eQq})^{(x)} \frac{\sqrt{6}}{4i^{(x)}(2i^{(x)}-1)} T^{(2)}(\hat{i}^{(x)},\hat{i}^{(x)}) \cdot C^{(2)}(\hat{r}^{(X)}).
\end{align}
where $eQq$ is the coupling constant, $\hat{i}^{(x)}$ is the rank-1 nuclear
spin operator for nucleus $x$, $C^{(2)}(\hat{r}^{(X)})$ is the rank-2 tensor with spherical
components that are Racah-normalized spherical harmonics,
$C_{2,q}(\hat{r}^{(X)})$, as a function of the spherical polar angles of the
molecular axis in the space-fixed frame, and
\begin{align}
T^{(k)}_q(\hat{A}^{(k_A)},\hat{B}^{(k_B)}) = \sum_{q_A,q_B} \hat{A}^{(k_A)}_{q_A} \hat{B}^{(k_B)}_{q_B} \langle k_A q_A k_B q_B | k q \rangle
\end{align}
is the $q$ spherical component of the rank-$k$ irreducible spherical tensor
product of $\hat{A}^{(k_A)}$ and $\hat{B}^{(k_B)}$, and $\langle j_1 m_1 j_2
m_2 | j m \rangle$ is a Clebsch-Gordan coefficient. The dot product of two
tensors is related to this rank-0 tensor product by
\begin{align}
\hat{A}^{(k)} \cdot \hat{B}^{(k)} = (-1)^k \sqrt{2k+1} T^{(0)}_0(\hat{A}^{(k)},\hat{B}^{(k)}),
\end{align}
and reduces to the usual dot product of two vectors for $k=1$.

Further hyperfine couplings include the spin-rotation coupling, $c_x \hat{i}^{(x)} \cdot \hat{n}$,
a dipolar term given by
\begin{align}
H^{(X)}_3 = -c_3\sqrt{6} T^{(2)}(\hat{i}^{(X_1)}, \hat{i}^{(X_2)}) \cdot C^{(2)}(\hat{R}),
\end{align}
and finally a term $c_4 \hat{i}^{(X_1)} \cdot \hat{i}^{(X_2)}$,
which arises due to electron-mediated dipolar interactions.

Application of a magnetic field, $\vec{B}$, induces the following Zeeman interactions
\begin{align}
\hat{H}^{(X)}_\mathrm{Zeeman} = - g_{\rm r} \mu_\mathrm{B} \hat{n} \cdot \vec{B} - g_1 \mu_\mathrm{N} \hat{i}^{(X_1)} \cdot \vec{B} - g_2 \mu_\mathrm{N} \hat{i}^{(X_2)} \cdot \vec{B}.
\end{align}
The three terms correspond to the interactions between the external magnetic field and the magnetic moments associated with the rotation of the molecule and the nuclear spins $1$ and $2$, respectively.
We do not include the interaction of these magnetic moments with the microwave magnetic field.

In this work, we consider RbCs and KCs molecules, where the $^{87}$Rb,
$^{39}$K, and $^{133}$Cs nuclei have spins 3/2, 3/2, and 7/2, respectively. The
hyperfine coupling constants are summarized in Table~\ref{tab:molconst}. We
also consider $^{40}$Ca$^{19}$F($^2\Sigma^+$) molecules in this work. In this
case, the $^{19}$F nucleus and the \emph{electrons} contribute 1/2 spin, while
$^{40}$Ca has zero nuclear spin. The couplings involving the electron spin have
a similar form as discussed above for the hyperfine interactions of $^1\Sigma$
molecules, but the coupling constants are typically much larger, as can be seen
in Table~\ref{tab:molconst}.

\setlength{\tabcolsep}{30pt}
\begin{table}
\caption{
Rotational constants, dipole moments and coupling constants for the fine and hyperfine structure of the \RbCs, \KCs, and \CaF\ molecules used in this work, from Refs.~\cite{aldegunde:17,aymar:05}.
\label{tab:molconst}}
\begin{tabular}{lrrr}
\hline\hline
 & $^{87}$Rb$^{133}$Cs & \KCs\ & $^{40}$Ca$^{19}$F \\
\hline
$b_\mathrm{rot}$ & 490.17 MHz & 1.0~GHz & 10.267 GHz \\
$\mu$ & 1.225 Debye & 1.92 Debye & 3.07 Debye \\
\\
$i^{(1)}$ & 3/2 & 3/2 & 1/2  \\
$i^{(2)}$ & 7/2 & 7/2 & 1/2  \\
$(eQq)^{(1)} $ & -809.29 kHz & -182 kHz & 0 \\
$(eQq)^{(2)} $ & 59.98 kHz & 75 kHz & 0 \\
$c_1$ &  98.4 Hz & 8.6 Hz & 39.659 MHz \\
$c_2$ & 194.1 Hz & 385.4 Hz & 29.07 kHz\\
$c_3$ & -192.4 Hz & 18 Hz & 13.37 MHz \\
$c_4$ & 17.3454 kHz & 1.1463 kHz & 122.556 MHz\\
& & \\
$g_{\rm r}$ & 0.0062 & 0 & 0 \\
$g_1$ & 1.8295 & 0.261 & -2.0023 $\mu_\mathrm{B}/\mu_\mathrm{N}$ \\
$g_2$ & 0.7331 & 0.738 & 0 \\
\hline\hline
\end{tabular}
\end{table}

\subsection{Dimer Hamiltonian}

The total Hamiltonian is given by
\begin{align}
\hat{H} = -\frac{\hbar^2}{2M} \frac{1}{R} \frac{d^2}{dR^2} R + \frac{\hbar^2 \hat{L}^2}{2M R^2} + \hat{H}^{(A)} + \hat{H}^{(B)} + \hat{V}(R).
\end{align}
Here, the reduced mass is $M$, the distance between the molecules is $R$, and
$\hat{L}$ is the dimensionless angular momentum operator associated with the end-over-end
rotation of the intermolecular axis, $\vec{R}$. The first term describes the
radial kinetic energy and the second the centrifugal kinetic energy. The third
and fourth terms correspond to the monomer Hamiltonian discussed above. These
describe the rotation of the molecules, which are treated as rigid rotors, as
well as the interaction with the microwave radiation and possible static
external fields, and all nuclear hyperfine interactions, where included. The
interaction between the two molecules is denoted by the interaction potential
$\hat{V}$. In this work, the inter-molecular potential is given by the
dipole-dipole interaction
\begin{align}
\hat{V}(R) = -\frac{\sqrt{6}}{4\pi\epsilon_0 R^3} T^{(2)}(\hat{\mu}^{(A)},\hat{\mu}^{(A)}) \cdot C^{(2)}(\hat{R})
\end{align}

\subsection{Basis sets}

We use completely uncoupled basis sets.
For monomer $X=A,B$, the basis set consists of products of rotational states, $|n_X m_{n_X}\rangle$,
\begin{align}
\langle \bm{r}_X | n_X m_{n_X} \rangle = \sqrt{\frac{2n+1}{4\pi}} C_{n,m_{n_X}}(\hat{r}_X),
\end{align}
angular momentum kets for (nuclear) spin states, if included, $| i_{X_1} m_{i_{X_1}} \rangle$,
and a photon state $|N\rangle$, where $N+N_0$ is the number of photons with polarization $\sigma$ and detuning $\Delta$,
and $N_0\gg1$ is the reference number of photons.
Thus, the basis functions for monomer $X$ take the form
\begin{align}
|n_X m_{n_X}\rangle|i_{X_1} m_{i_{X_1}}\rangle|i_{X_2} m_{i_{X_2}}\rangle|N\rangle.
\end{align}

For the dimer, we introduce an angular momentum state, $|L M_L\rangle$, that
describes the end-over-end rotation of the intermolecular axis,
\begin{align}
\langle \vec{R} | L M_L \rangle = \sqrt{\frac{2L+1}{4\pi}} C_{L,m_{L}}(\hat{R}).
\end{align}
The dimer basis functions then take the form \cite{hanna:10,owens:16,owens:17}
\begin{align}
|n_A m_{n_A}\rangle|i_{A_1} m_{i_{A_1}}\rangle|i_{A_2} m_{i_{A_2}}\rangle|n_B m_{n_B}\rangle|i_{B_1} m_{i_{B_1}}\rangle|i_{B_2} m_{i_{B_2}}\rangle|L M_L\rangle|N\rangle.
\end{align}
Only even values of $L$ are included: The only interaction that couples states
with different $L$ is the dipole-dipole interaction, and this conserves the
parity of $L$.

The dimer basis set includes only functions with a single value of
$\mathcal{M}=m_{n_A}+m_{i_{A_1}}+m_{i_{A_2}}+m_{n_B}+m_{i_{B_1}}+m_{i_{B_2}}+M_L
+ \sigma N$, which is set by $\mathcal{M}$ for the relevant initial state with
$M_L=L=0$. Different values of $\mathcal{M}$ are not coupled. Furthermore, the
basis functions are adapted to permutation symmetry by acting with
$1+\hat{\mathcal{P}}_\mathrm{ab}$, where the action of the permutation operator
is
\begin{align}
\hat{\mathcal{P}}_\mathrm{ab} |n_A m_{n_A}\rangle|i_{A_1} m_{i_{A_1}}\rangle|i_{A_2} m_{i_{A_2}}\rangle|n_B m_{n_B}\rangle|i_{B_1} m_{i_{B_1}}\rangle|i_{B_2} m_{i_{B_2}}\rangle|L M_L\rangle|N\rangle = \nonumber \\
(-1)^L |n_B m_{n_B}\rangle|i_{B_1} m_{i_{B_1}}\rangle|i_{B_2} m_{i_{B_2}}\rangle |n_A m_{n_A}\rangle|i_{A_1} m_{i_{A_1}}\rangle|i_{A_2} m_{i_{A_2}} |L M_L\rangle|N\rangle.
\end{align}

We have carried out convergence tests with respect to the truncation of the
basis set. The calculated probabilities for RSR and microwave-induced loss
rates are converged to approximately $1$~$\%$. This requires inclusion of
functions with $N=0$, $-1$, and $-2$ photons, rotational states up to
$n_X^\mathrm{max}=1$, and partial waves up to $L^\mathrm{max}=6$. Where fine
and/or hyperfine interactions are included, the range of $m_i$ functions needed
for convergence is limited to those that differ by at most two quanta from
$m_i$ that contribute to the initial state. In the case of CaF+CaF, this means
that all fine and hyperfine states are included. For calculations including
static electric fields, rotational states up to $n_X^\mathrm{max}=2$ were
included to obtain converged loss rates, and $L^\mathrm{max}=20$ is needed in
the calculation of elastic cross sections.

\subsection{Cross sections and rates}

We perform coupled-channels scattering calculations using the renormalized
Numerov method. We impose capture boundary conditions using the method of
Janssen \emph{et al.}\ \cite{janssen:13}. This yields an $S$-matrix for the
combined set of short-range and long-range product channels. Cross sections can
be computed from the $T$-matrix, $\bm{T}=\bm{1}-\bm{S}$, as follows
\begin{align}
\sigma_{i\rightarrow f} = \frac{\pi}{k^2} \sum_{L,M_L,L',M_L'} \left| T^{(\mathrm{LR})}_{f,L',M_L';i,L,M_L} \right|^2,
\end{align}
where $(\mathrm{LR})$ indicates the long-range part, and the matrix element
$T^{(\mathrm{LR})}_{f,L',M_L';i,L,M_L}$ refers to the initial and final states
$i$, $f$, with relative angular momentum quantum numbers $L$, $M_L$ and $L'$,
$m'_L$, respectively.
We define three cross sections: the elastic cross section, $\sigma_\mathrm{el}
= \sigma_{i\rightarrow i}$, the inelastic cross section (microwave-induced
here), $\sigma_{\mathrm{inel}} = \sum_{f \neq i} \sigma_{i\rightarrow f}$, and
the RSR (capture) cross section
\begin{align}
\sigma_\mathrm{RSR} = \frac{\pi}{k^2} \sum_{r,L,M_L} \left| T^{(\mathrm{SR})}_{r;i,L,M_L} \right|^2.
\end{align}
Here, $(\mathrm{SR})$ denotes the short-range capture part and the sum over $r$
extends over all adiabatic channels that are classically allowed at the capture
radius.
In the threshold regime, the elastic cross section is independent
of the collision energy, whereas the cross sections for exoenergetic inelastic
collisions and short-range capture scale with $E^{-1/2}$.

Thermal rate coefficients can be calculated by averaging the cross sections over a Maxwell-Boltzmann distribution,
\begin{align}
k^{(2)} = \sqrt{\frac{8 k_{\rm B}T}{\pi\mu}} \frac{1}{(k_{\rm B}T)^2} \int_0^\infty \sigma(E) \exp\left(-\frac{E}{kT}\right) E dE.
\end{align}
This results in the rate coefficient
\begin{align}
k^{(2)} &= \sqrt{\frac{2}{\mu}} \left[E^{1/2}\sigma(E)\right],
\end{align}
for cross sections that scale with $E^{-1/2}$.
This allows calculation of temperature-independent rate coefficients in the threshold regime.
For microwave-induced loss this corresponds to an actual loss rate,
but it may be that not all flux that reaches short range is lost.
In the present work, all loss rates were calculated for a collision energy of 1 $\mu$K $\times\ k_{\rm B}$.

We also define RSR and microwave-induced loss \emph{probabilities}, as the sum
of squares of the relevant $S$-matrix elements
\begin{align}
P_\mathrm{RSR} &= \sum_{r} \left| S^{(\mathrm{SR})}_{r;i,0,0} \right|^2, \nonumber \\
P_\mathrm{inel} &= \sum_{f\neq i, L',M_L'} \left| S^{(\mathrm{LR})}_{f,L',M_L';i,0,0} \right|^2, \nonumber \\
P_\mathrm{shielding} &= \sum_{L',M_L'} \left| S^{(\mathrm{LR})}_{i,L',M_L';i,0,0} \right|^2 = 1-P_\mathrm{short~range}-P_\mathrm{trap~loss}.
\end{align}
The RSR probability provides a measure of the reduction of flux reaching short
intermolecular distances.

\subsection{Validation}

\begin{figure}
\begin{center}
\includegraphics[width=\figwidth,clip]{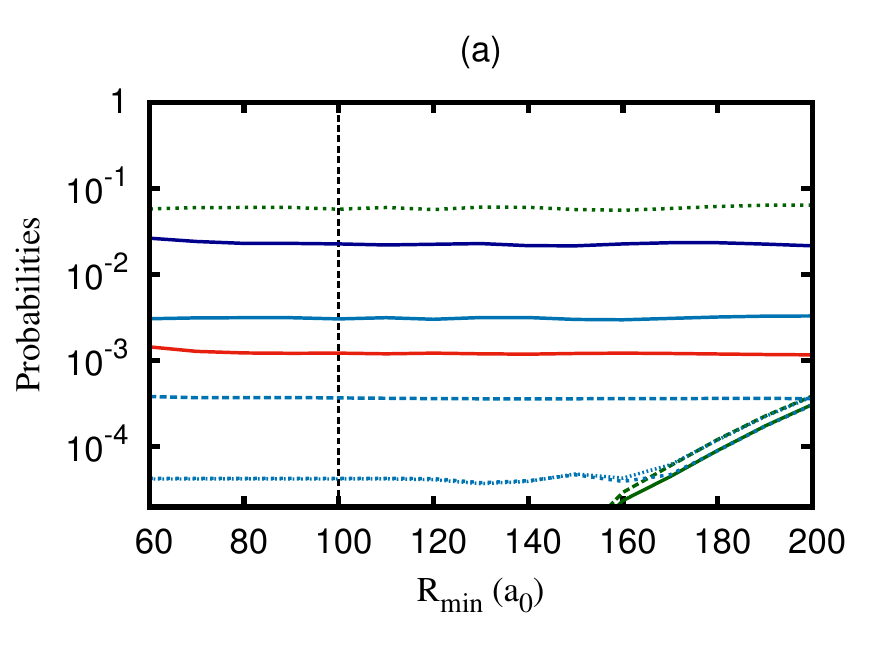}
\includegraphics[width=\figwidth,clip]{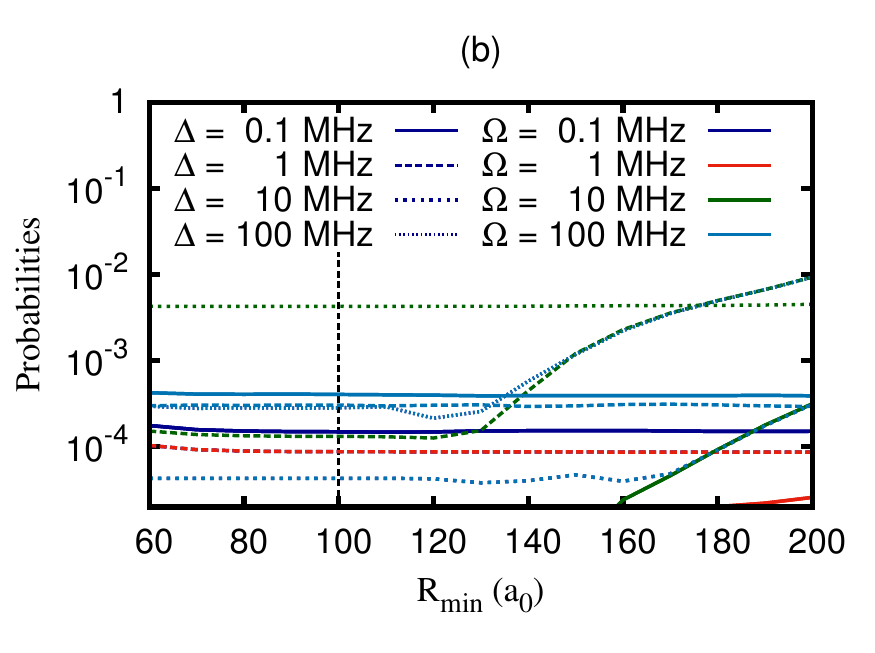}
\caption{ \label{fig:Rmin}
Panels (a) and (b) show probabilities for RSR and microwave-induced loss for RbCs neglecting hyperfine interactions, as a function of $R_\mathrm{min}$ for various $\Omega$ and $\Delta$.
The vertical dashed line indicates the value $R_\mathrm{min}=100~a_0$ used in our calculations.}
\end{center}
\end{figure}

The results of our scattering calculations are converged with respect to
basis-set truncation as described above. Here, we investigate their sensitivity
to the distance where the short-range capture boundary conditions are imposed.
Figure~\ref{fig:Rmin} shows the dependence of probabilities for RSR and
microwave-induced loss on the capture radius $R_\mathrm{min}$ for various
$\Omega$ and $\Delta$. The calculations elsewhere all use the converged value
$R_{\rm min}=100~a_0$.

Additional tests that confirm correct implementation of the capture boundary
conditions include:
\begin{itemize}
\item For an $R^{-6}$ potential, the loss rate due to RSR agrees with the
    universal loss rates obtained analytically by Idziaszek and Julienne
    \cite{Idziaszek:PRL:2010}.
\item For a capture radius well inside the classically forbidden region,
    our code reproduces the results of ``standard'' scattering calculations
\end{itemize}
We have also verified that we obtain identical loss rates for linear
polarization parallel to the quantization axis and perpendicular to it. These
are implemented as $\pi$ polarization and as linear combinations of $\sigma^+$
and $\sigma^-$ polarizations, respectively. The two calculations are very
different because the projection $\mathcal{M}$ along $z$ is not conserved for
linear polarization in the $xy$ plane \cite{hanna:10}.

\section{Numerical Results}
\subsection{{RbCs}}
\subsubsection{Hyperfine free}

\begin{figure}
\begin{center}
\includegraphics[width=\textwidth,clip]{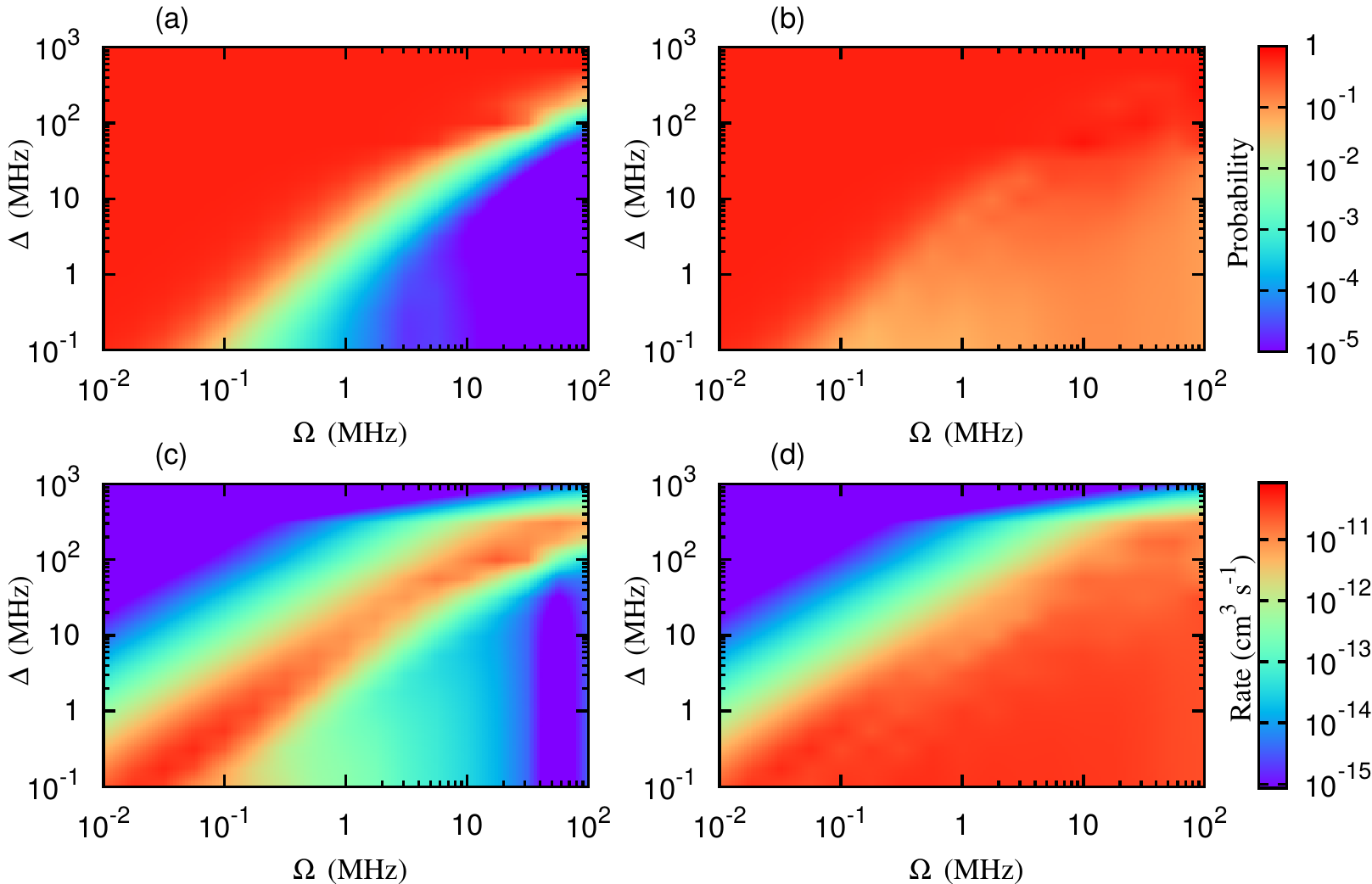}
\caption{ \label{fig:polariz}
Loss probabilities and rates for RbCs+RbCs collisions without static fields or
hyperfine interactions. Panels (a) and (b) show the probability for reaching
short range for circular and linear polarization, respectively. Panels (c) and
(d) show the microwave-induced loss rate for circular and linear polarization.
}
\end{center}
\end{figure}

Figure~\ref{fig:polariz} shows the RSR and microwave-induced loss probabilities
and rates for both circularly and linearly polarized microwaves, as a function
of the detuning, $\Delta$, and the Rabi frequency, $\Omega$, for RbCs+RbCs
collisions without static fields or hyperfine interactions. Shielding is
ineffective for linear polarization. This has also been observed in optical
blue shielding of ultracold atoms \cite{weiner:99}, and can be understood in
terms of missing Q-branch couplings between ground and excited channels with
equal total angular momentum \cite{napolitano:97}. It is caused by vanishing
coupling between the ground and repulsive excited states when the molecules
collide along the direction of linear polarization.

\begin{figure}
\begin{center}
\includegraphics[width=\figwidth,clip]{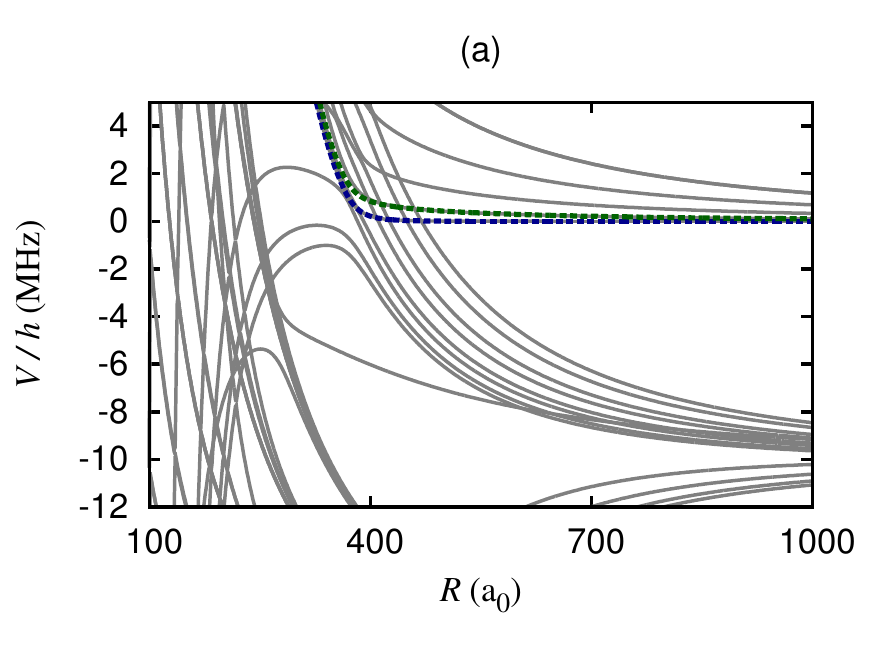}
\includegraphics[width=\figwidth,clip]{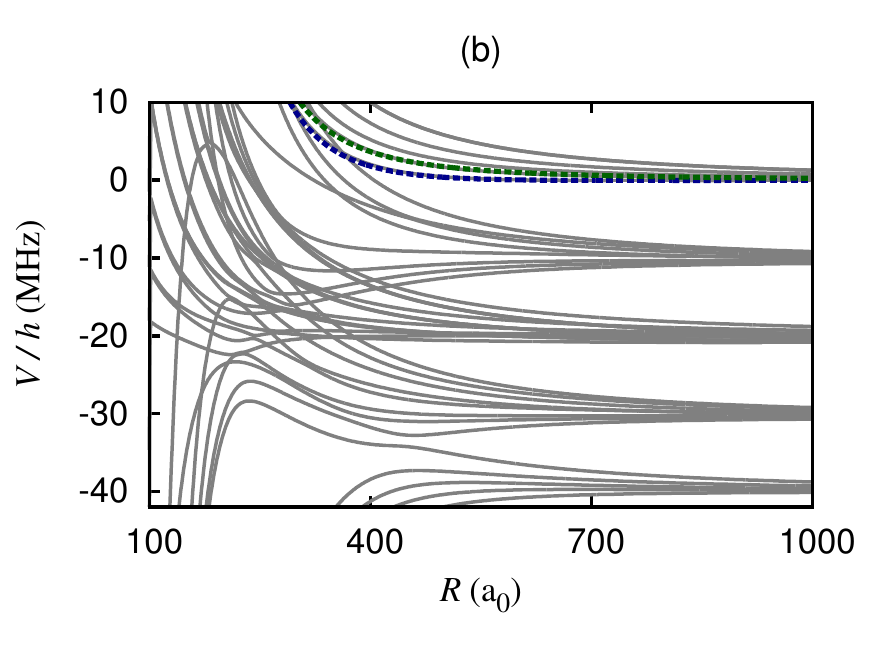}
\includegraphics[width=\figwidth,clip]{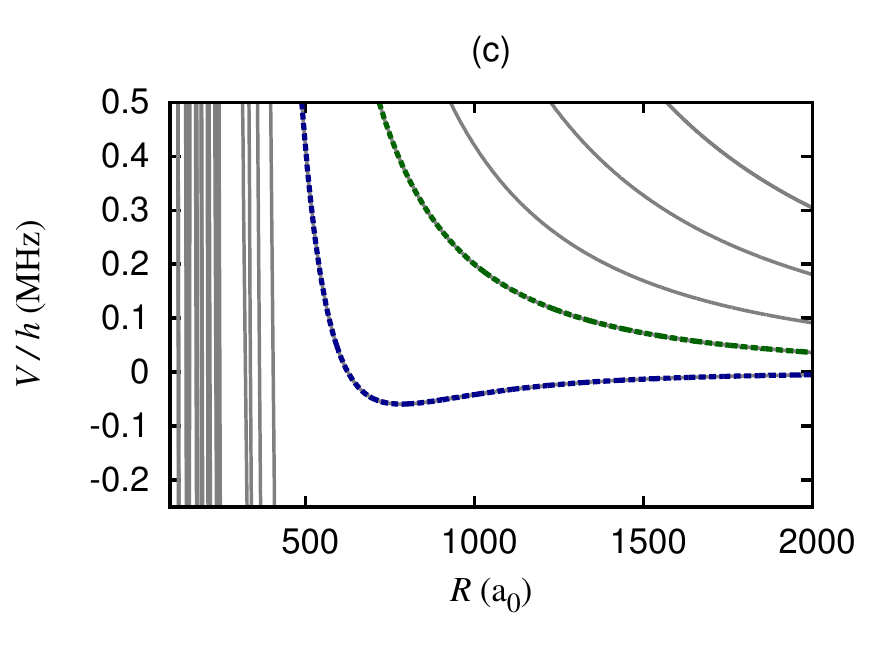}
\caption{ \label{fig:RbCs_pots}
Adiabatic potential curves for RbCs+RbCs with $L^\mathrm{max}=8$. Panel (a)
corresponds to $\Delta=10$~MHz and $\Omega=1$~MHz; the remaining panels
correspond to $\Delta=1$~MHz and $\Omega=10$~MHz. The adiabats are shown by
gray solid lines, whereas to guide the eye the initial $s$ and $d$-wave
channels are marked by superimposed blue and green dotted lines, respectively.
Panel (c) offers an expanded view of the curves in panel (b). }
\end{center}
\end{figure}

\subsubsection{Adiabatic potential curves}

Figure~\ref{fig:RbCs_pots} shows adiabatic potential energy curves, defined as
the eigenvalues at fixed intermolecular distance, $R$, of the Hamiltonian
without radial kinetic energy. These are shown for RbCs+RbCs with the
partial-wave quantum number truncated at $L^\mathrm{max}=8$. They differ from
the schematic representation of Fig.~1 of the main text because of the
inclusion of the end-over-end rotation $L$. Panel (a) shows potentials for
$\Delta=10$~MHz and $\Omega=1$~MHz, where the blue and green dashed lines
indicate the initial $s$ and $d$-wave, respectively. The potentials in this
case resemble the schematic Fig.~1(a) of the main text, where the $R^{-6}$
potential of the bare ground state is almost flat outside the Condon
point, and crosses $R^{-3}$ repulsive potentials coming up from $-\hbar\Delta$.
Panel (b) shows potentials for $\Delta=1$~MHz and $\Omega=10$~MHz, which
provide effective shielding of molecular collisions. These potentials
correspond more closely to schematic Fig.~1(b) of the main text. An expanded
view is shown in panel (c), which demonstrates that there is no clear avoided
crossing. The repulsive potentials of the top field-dressed manifold result
from dipole-dipole couplings between the different manifolds.

\subsubsection{Hyperfine structure}

\begin{figure}
\begin{center}
\includegraphics[width=\figwidth,clip]{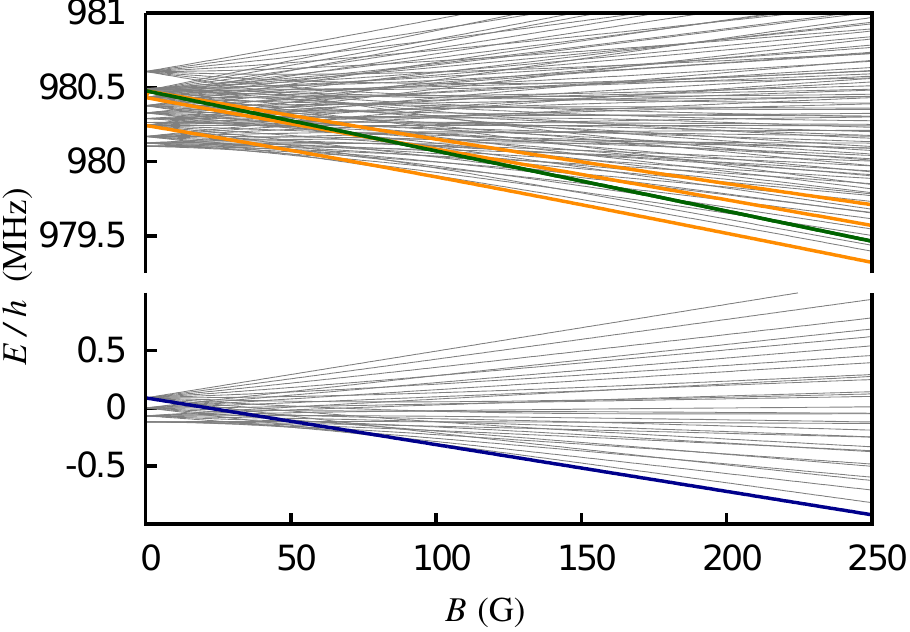}
\caption{ \label{fig:RbCs_hyperfine}
Hyperfine structure of \RbCs\ as a function of magnetic field. Marked in blue
and green are the spin-stretched rotational ground and excited states with
quantum numbers $f=m_f=5$ and $f=m_f=6$, respectively. Marked in orange are
rotationally excited states with $m_f=5$. }
\end{center}
\end{figure}

Figure~\ref{fig:RbCs_hyperfine} shows the hyperfine states of \RbCs\ as a
function of the magnetic field strength \cite{gregory:16}. The molecules
are initially in their spin-stretched rotational ground state, $f=m_f=5$, which
is shown in blue. Circularly polarized microwaves provide coupling to the
$f=m_f=6$ rotationally excited state, shown in green. Orange lines mark
rotationally excited states with $m_f=5$.

\subsection{KCs}

\begin{figure}
\begin{center}
\includegraphics[width=\figwidth,clip]{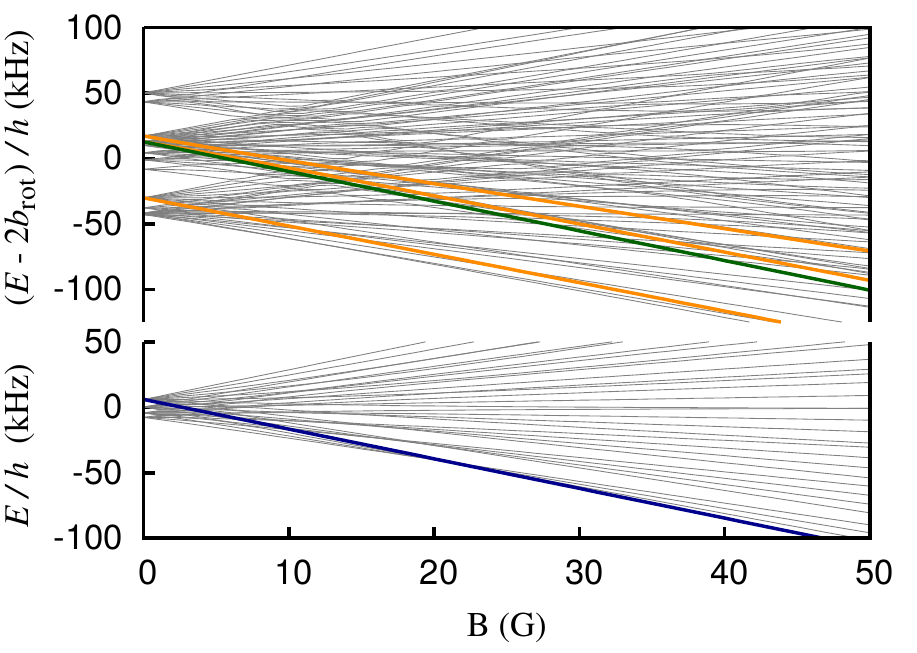}
\caption{ \label{fig:KCs_hyperfine}
Hyperfine structure of \KCs\ as a function of magnetic field. Marked in blue
and green are the spin-stretched rotational ground and excited states with
quantum numbers $f=m_f=5$ and $f=m_f=6$, respectively. Marked in orange are
rotationally excited states with $m_f=5$.  }
\end{center}
\end{figure}

Here, we present further numerical results for \KCs\ molecules.
Figure~\ref{fig:KCs_hyperfine}(a) shows the hyperfine states as a function of
magnetic field. This uses the same color-coding as for RbCs, namely blue and
green lines correspond to spin-stretched rotational ground and excited states,
and orange lines correspond to $m_f=5$ excited states. The hyperfine splittings
of the rotational ground state, that are determined by the $c_4 \hat{i}^{(1)} \cdot \hat{i}^{(2)}$ term, are much smaller for \KCs\ than they are for \RbCs.
For the rotationally excited state, interactions involving the nuclear quadrupole
moment dominate and cause splittings in the order of 100~kHz and 500~kHz for KCs and RbCs, respectively.

\begin{figure}
\begin{center}
\includegraphics[width=\textwidth,clip]{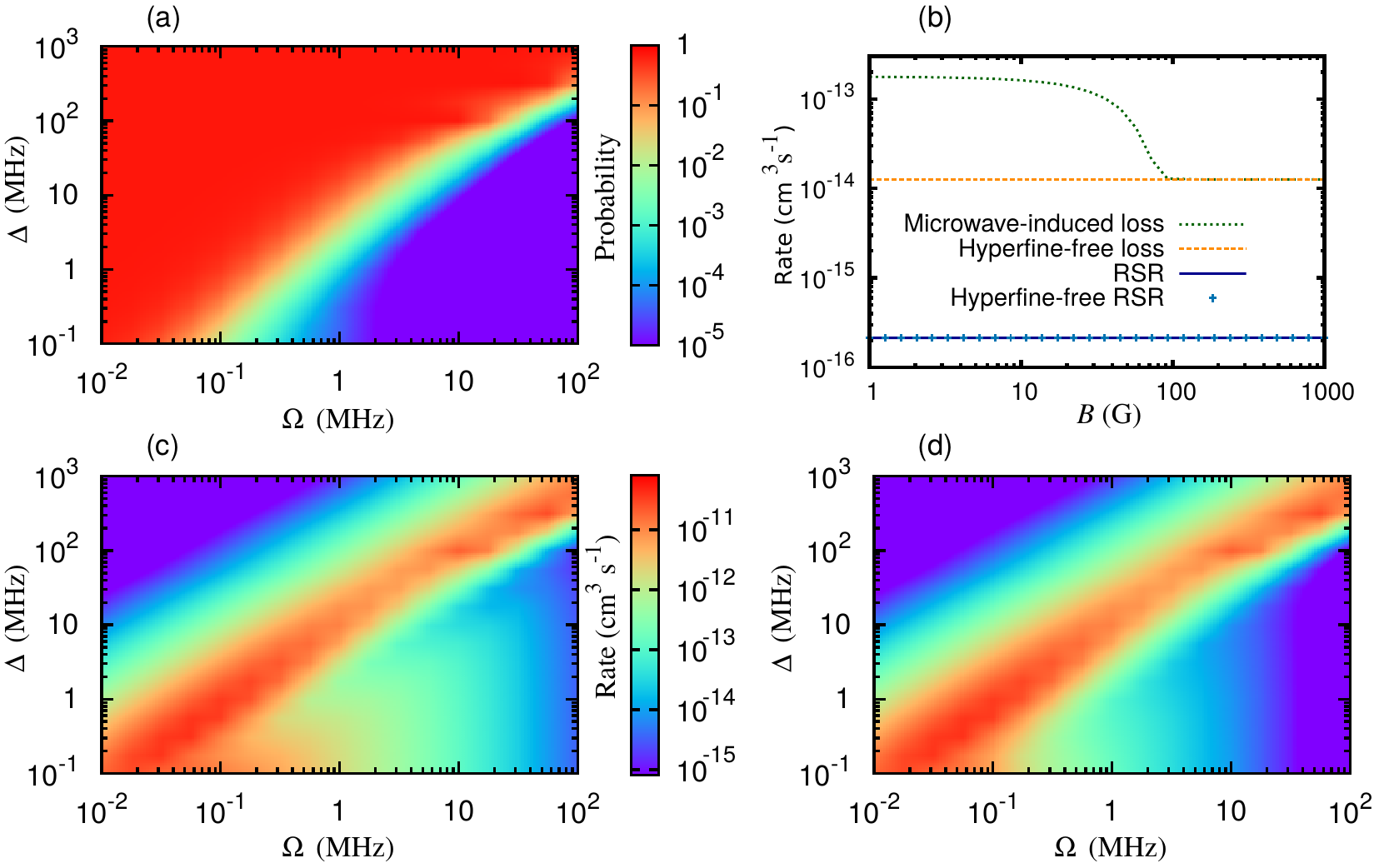}
\caption{ \label{fig:KCs_shielding}
Loss probabilities and rates for KCs+KCs collisions including hyperfine
interactions. Panel (a) shows the probability for RSR in the absence of a
magnetic field. Panel (b) shows the dependence of the RSR and microwave-induced
loss rates on magnetic field for fixed $\Omega=10$~MHz and $\Delta=1$~MHz.
Panel (c) and (d) show the microwave-induced loss rate in the absence of a
magnetic field, and for a magnetic field of 200~G, respectively.}
\end{center}
\end{figure}

Figure~\ref{fig:KCs_shielding} shows results for shielding of KCs molecules.
Panels (a) and (c) show the probability of RSR and microwave-induced loss rates, respectively,
without static electric or magnetic fields. Microwave-induced loss for large
$\Omega$ is somewhat larger than in the hyperfine-free case, but not by nearly
as much as for \RbCs. The reason is the weaker coupling of the nuclear spins to
the molecular axis, such that the nuclear spins retain their space-fixed
quantization even for non-spin-stretched states. The increased loss that does
occur can be suppressed using magnetic fields. This is shown in panel (d) for a
magnetic field of 200~G. Panel (b) shows the field-dependence of the
microwave-induced and RSR rate at fixed $\Delta=1$~MHz and $\Omega=10$~MHz. For
fields above 100~G, we reach a regime where the Zeeman interaction dominates
and the individual spin projections and $m_n$ become nearly good quantum
numbers. The spins then play the role of spectators, and the losses are
suppressed to the hyperfine-free level.

\subsection{CaF}

\begin{figure}
\begin{center}
\includegraphics[width=\figwidth,clip]{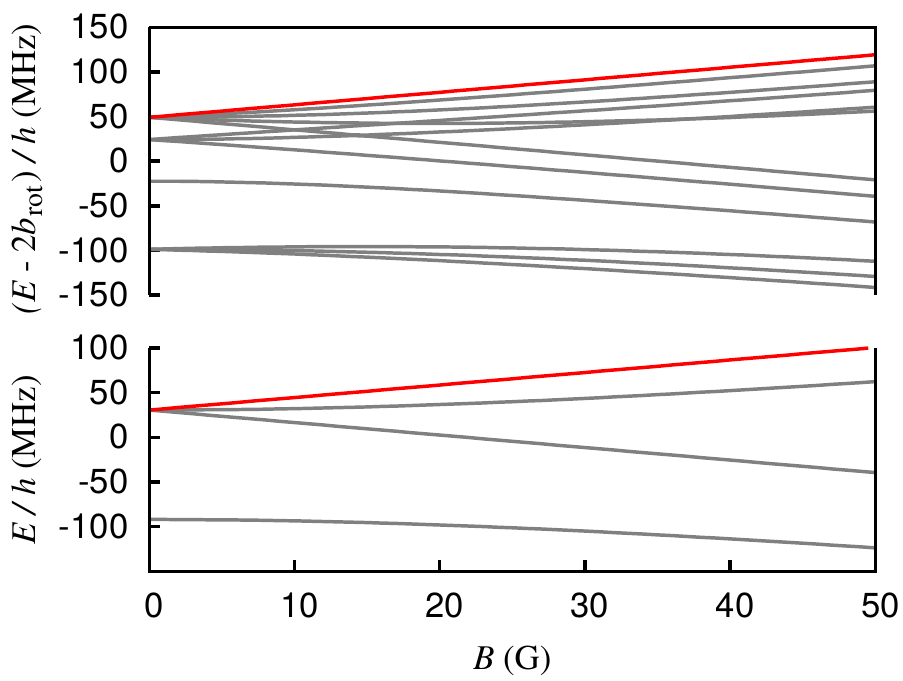}
\caption{ \label{fig:CaF_hyperfine}
Hyperfine structure of \CaF\ as a function of magnetic field strength.
Red lines indicate the spin-stretched rotational ground and excited states.
}
\end{center}
\end{figure}

We also consider shielding of CaF($^2\Sigma$) molecules. The fine and hyperfine states
of \CaF\ are shown as a function of the magnetic field in
Fig.~\ref{fig:CaF_hyperfine} \cite{Blackmore:2018}. Red lines denote the
spin-stretched ground and excited states. Molecules in the spin-stretched
states can be trapped magnetically, and the magnetically insensitive transition
between them is of interest for coherent spectroscopy in such magnetic traps
\cite{Blackmore:2018}. The couplings involving the electron spin are much
stronger than hyperfine couplings in alkali-metal dimers. This results in much
larger splittings at zero magnetic field and a stronger coupling of the spins
to the molecular axis.

\begin{figure}
\begin{center}
\includegraphics[width=\textwidth,clip]{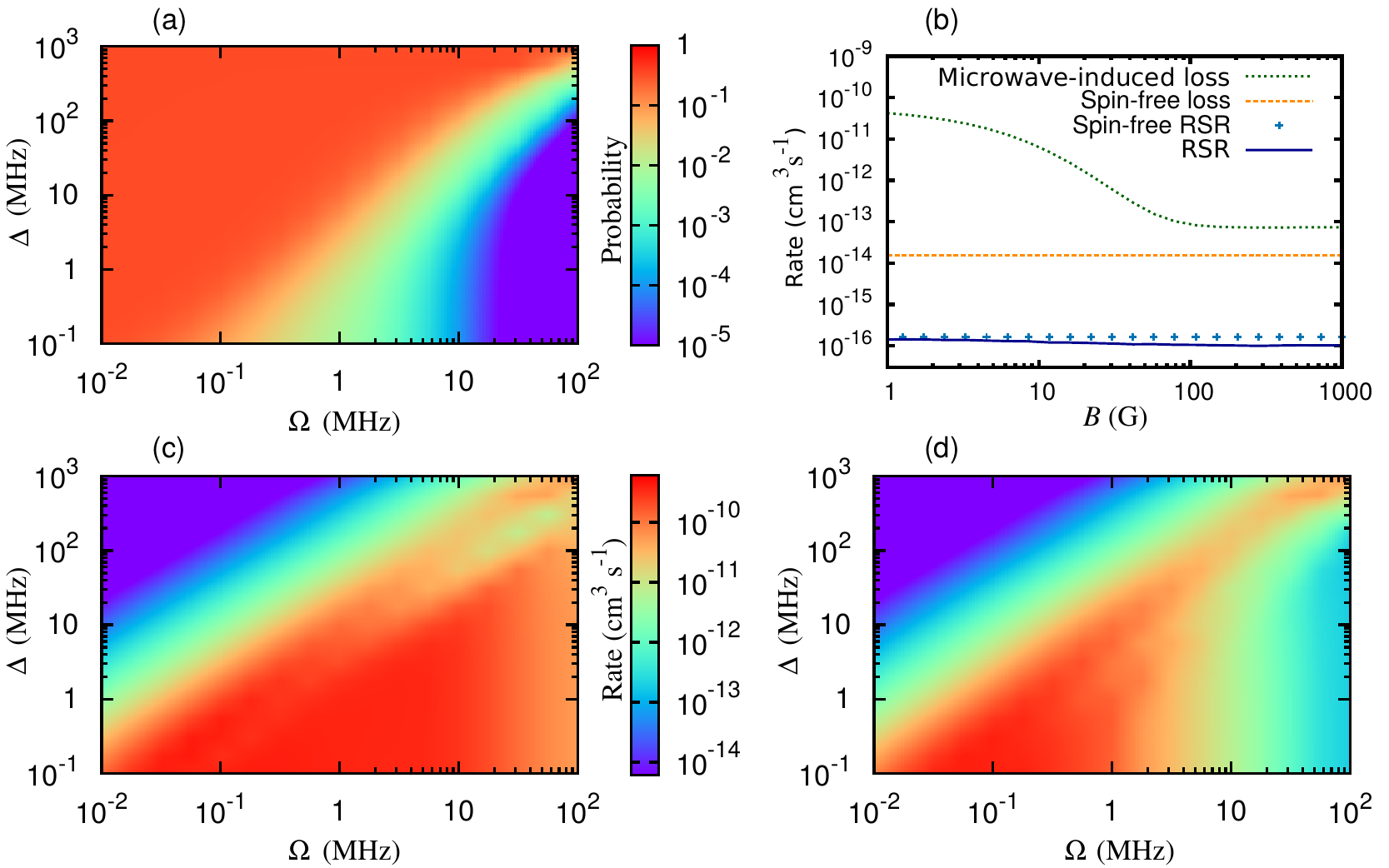}
\caption{ \label{fig:CaF_shielding}
Loss probabilities and rates for CaF+CaF collisions including fine and hyperfine
interactions. Panel (a) shows the probability for RSR in the absence of a
magnetic field. Panel (b) shows the dependence of the RSR and microwave-induced
loss rates on magnetic field for fixed $\Omega=100$~MHz and $\Delta=1$~MHz.
Panel (c) and (d) show the microwave-induced loss rate in the absence of a
magnetic field, and for a magnetic field of 50~G, respectively.}
\end{center}
\end{figure}

Results for shielding of \CaF\ molecules are shown in
Fig.~\ref{fig:CaF_shielding}. Panels~(a) and (c) show probabilities for RSR
and microwave-induced loss rates at zero magnetic field, respectively. The microwave-induced
loss rates are much larger than for bialkali molecules. This loss can again be
suppressed by applying a magnetic field, as is shown in panel~(d) for a field of 50~G. The dependence of the loss on magnetic field is
shown in panel (b) for fixed $\Delta=1$~MHz and $\Omega=100$~MHz.
Microwave-induced loss is suppressed as the spins uncouple from the molecular
axis and quantize along the field. However, even in the high-field limit, the
rates differ from the results of a spin-free calculation because the
spin-rotation coupling constant $\gamma = c_1 \simeq 40$~MHz is not negligible
compared to $\Omega$.

\section{Adiabatically switching on microwaves}

Microwave shielding requires preparation of the molecules in the upper state of
the molecule-field Hamiltonian. We consider here the time needed to switch on a
microwave field without significant nonadiabatic losses. We use the basis set
$\{ |0 0 0\rangle, |1,1,-1\rangle \}$. In this basis set, the Hamiltonian is
represented by the 2$\times$2 matrix
\begin{align}
\mathbf{H} = \left[ {\begin{array}{cc} 0 & \hbar\Omega \\
\hbar\Omega & -\hbar\Delta \\
\end{array}} \right].
\end{align}
The upper (lower) adiabatic state in this basis is given by the vector
$[\cos\varphi\ \sin\varphi]^T$ ($[-\sin\varphi\ \cos\varphi]^T$), where the
mixing angle is given by
\begin{align}
\varphi = \frac{1}{2} \tan^{-1}\left( \frac{2\Omega}{\Delta} \right).
\end{align}
For $\Omega\ll\Delta$ the mixing angle is small, $\varphi\approx
\Omega/\Delta$. As a result, even if the microwaves are turned on
instantaneously, most molecules will be found in the upper adiabatic state.
This is typically the case for blue shielding of atoms, where shielding is
effective for $\Omega < \Delta$ \cite{weiner:99}.
For polar molecules, however, we have seen that efficient shielding often
requires $\Omega$ comparable to $\Delta$, or even $\Omega\gg\Delta$.

We take the Rabi coupling to be the following function of time,
\begin{align}
\Omega(t) = \begin{cases}
0 & t<0, \\
\Omega_\infty \sqrt{\frac{t}{\tau}} & 0<t<\tau,\\
\Omega_\infty & t>\tau.
\end{cases}
\end{align}
Hence, the microwaves are switched on, from zero to the desired intensity,
$\Omega_\infty$, between times $t=0$ and $t=\tau$. The square-root time
dependence is chosen such that microwave \emph{intensity} is linear in time.
The time-dependent wave function is given by
\begin{align}
|\psi(t)\rangle =
\sum_{i} a_i \exp\left(-i\epsilon_i t/\hbar\right) |i\rangle,
\end{align}
where the instantaneous eigenstates satisfy
\begin{align}
\hat{H}(t) | i \rangle = \epsilon_i(t) |i\rangle.
\end{align}
The expansion coefficients $a_i$ satisfy
\begin{align}
\frac{d a_i}{dt} =
- \sum_{j \neq i} \langle i | \frac{d}{dt} | j \rangle
\exp\left[\frac{i}{\hbar} \left(\epsilon_i-\epsilon_j\right) t \right] a_j =
\sum_{j \neq i} \frac{1}{\epsilon_i-\epsilon_j} \langle i | \frac{d \hat{H}}{dt} | j \rangle
\exp\left[\frac{i}{\hbar} \left(\epsilon_i-\epsilon_j\right) t \right] a_j.
\end{align}
We propagate these amplitudes, starting in the microwave-free ground state at
$t=0$ and propagate numerically to $t=\tau$ using an ODE solver \cite{odepack}.

For $\Omega=10$~MHz and $\Delta=1$~MHz, a switching time of $\tau=1$~ms
prepares 99~\% of the molecules in the upper adiabatic state for the linear
intensity ramp discussed above. If the intensity is ramped quadratically,
corresponding to a linear ramp of the Rabi frequency, the switching time $\tau$
can be reduced to 10~$\mu$s for the same final intensity. The advantage of the
quadratic intensity ramp is that the switch is slower at early times, where the
adiabatic states are still close in energy.
These switching times apply to any molecule,
but do not include the effect of static fields or hyperfine interactions.

\end{document}